\newif\ifabstract
\abstracttrue
 \abstractfalse 
\newif\iffull
\ifabstract \fullfalse \else \fulltrue \fi

\documentclass[11pt]{article}
\usepackage{rotating}
\usepackage{multirow}
\usepackage{amsfonts}
\usepackage{amssymb}
\usepackage{amstext}
\usepackage{amsmath}
\usepackage{xspace}
\usepackage{theorem}
\usepackage{graphicx}
\usepackage{url}
\usepackage{graphics}
\usepackage{colordvi}
\usepackage{colordvi}
\usepackage{subfigure}

\advance \oddsidemargin by -0.8in
\newcommand{\myparskip}{3pt}
\parskip \myparskip

\begin{document}

\title{Software Quality Metrics for Geant4: An Initial Assessment\footnote{Presented at ANS RPSD 2014 - 18th Topical Meeting of the Radiation Protection \& Shielding Division of ANS, Knoxville, TN, September 14-18, 2014; Included on CD-ROM, American Nuclear Society, LaGrange Park, IL (2014)}}
\author{Elisabetta Ronchieri\thanks{INFN CNAF at Bologna. Email: {\tt elisabetta.ronchieri@cnaf.infn.it}}\and Maria Grazia Pia\thanks{INFN at Genoa. Email: {\tt maria.grazia.pia@ge.infn.it}} \and Francesco Giacomini\thanks{INFN CNAF at Bologna. Email: {\tt francesco.giacomini@cnaf.infn.it}}}

\begin{titlepage}
\maketitle

\thispagestyle{empty}

\begin{abstract}
In the context of critical applications, such as shielding and radiation protection, ensuring the quality of simulation software they depend on is of utmost importance. The assessment of simulation software quality is important not only to determine its adoption in experimental applications, but also to guarantee reproducibility of outcome over time. 

In this study, we present initial results from an ongoing analysis of Geant4 code based on established software metrics. The analysis evaluates the current status of the code to quantify its characteristics with respect to documented quality standards; further assessments concern evolutions over a series of release distributions. We describe the selected metrics that quantify software attributes ranging from code complexity to maintainability, and highlight what metrics are most effective at evaluating radiation transport software quality. The quantitative assessment of the software is initially focused on a set of Geant4 packages, which play a key role in a wide range of experimental applications and are representative of different software development processes. We provide an interpretation of the data resulting from measurements on the selected Geant4 packages, and discuss methods to improve them. 

This work can be used as a baseline for evaluating correlations between software quality embedded in the Geant4 development process and simulation observables produced by Geant4-based applications. The result provide constructive guidance both to improve key software tools, such as Geant4, and to estimate their contribution in risk analyses concerning sensitive applications.
\end{abstract}

\end{titlepage}

\section{Introduction}\label{sec: intro}

In the context of critical applications, such as shielding and radiation protection, ensuring the quality of simulation software they depend on is of utmost importance. The assessment of simulation software quality is crucial both to determine its adoption in experimental applications and to guarantee reproducibility of outcome over time. Geant4 \cite{agostinelli, allison} is a simulation system used in a wide range of experimental applications, some of which, such as radiation protection and biomedical studies, may involve critical use cases and would especially profit from an objective assessment of the software quality.

In this study, we present the first results from an ongoing analysis of Geant4 code based on established software metrics from the perspective of software engineering. The analysis evaluates the current status of the code to quantify its characteristics with respect to documented quality standards; further appraisals concern evolutions over a series of release distributions. The quantitative assessment of the software is initially focused on a subset of Geant4 packages, which play a key role in scientific applications and are representative of different software development processes. We describe the chosen software metrics \cite{kan} that map the entities of a software system to numeric values through mathematical definitions, quantifying software attributes ranging from functionality to maintainability. To measure such metrics, we exploited a set of software metrics tools, i.e. programs that implement a set of software metrics definitions. They allow software engineers to assess a software system by extracting the required attributes from the software and providing the corresponding metric values. To correlate software metrics values in a well-defined way, we used software quality standard models, such as the ISO/IEC 9126 standard \cite{isoiec1}, which aggregate numerical values to aid quality analysis and assessment.

With this work we aim to create a baseline for evaluating correlations between software quality embedded in the Geant4 development process and simulation observables produced by Geant4-based applications. Furthermore, we intend to highlight what metrics are most effective at evaluating radiation transport software quality. We are confident that the result will provide constructive guidance both to improve key software tools, such as Geant4, and to estimate their contribution in risk analysis concerning sensitive applications.

The remainder of this paper is structured as follows. We first discuss the approach adopted to fulfill our assessment (see Section \ref{sec: procedure}). Then, we present the setup of our experiments (see Section \ref{sec: exp}) and the way we collected data (see Section \ref{sec: dc}). We provide an initial interpretation of a subset of data resulting from measurements made on the selected Geant4 packages (see Section \ref{sec: de}). As the analysis is still ongoing at the time of submitting this abstract, some results will be documented in the final paper and in the conference presentation.

\section{Procedure Description}\label{sec: procedure}

The metrics we chose are defined in the literature. They are often complex, favoring different interpretations from those originally suggested and opening up to various implementations in metrics tools that produce incomparable values. Furthermore, physics research software is insufficiently represented in the software engineering literature, lacking appropriate metrics to quantify its quality. Therefore, in this first assessment, we decided to consider a wide set of software metrics tools and client analysis tools to gather as many measurements as possible and evaluate their role in the peculiar context of scientific software.

In the following five steps, we describe the procedure adopted to build our experiment:
\begin{enumerate}
\item We considered metrics that appeared interesting to our study. They mainly provide information about program size, code distribution, flow complexity and object-orientation, as described in the existing metrics suites \cite{chidamber, hyatt, abreu}. 
\item We identified the tools to collect metrics based on several criteria: a search on the internet by using standard search engines and specifying straightforward terms; a careful reading of references from related works; an evaluation of legal aspects; a sharp look at the supported programming languages. The selected tools are available without legal restrictions and able to analyze the same software with common metrics. 
\item The actual volume of Geant4 induced us to analyze a subset of Geant4 code in this initial study. As the selected Geant4 code is written in C++, the selected tools must be able to analyze at least the C++ programming language. 
\item The existing literature does not identify univocally an approach that determines the goodness of metrics values. We used them to quantify software attributes, which would reveal opportunities for possible improvement. With respect to software quality, we used the ISO 9126 standard \cite{isoiec2,isoiec3},
which defines six software characteristics or factors, each subdivided in sub-characteristics or criteria. 
\item We considered software quality models that assess software characteristics coming from the ISO 9126 standard. In this phase of our study, we mainly considered maintainability, the software factor that allows one to estimate how well software can improve in terms of correctness, adaptation and perfection \cite{ieee}. Therefore the selected models measure at least the aforementioned characteristic of the system.
\end{enumerate}

\section{Experimental Setup}\label{sec: exp}

We performed the experiments on a standard PC with the Ubuntu operating system version 13.10 satisfying the minimum requirements for all software metrics tools. We accomplished all measurements on this computer and saved data for further analysis.

\paragraph{Geant4 Package Selection}
Geant4 is a software system that simulates the passage of particles through matter. Geant4 is written in the C++ language and encompasses several packages, each one responsible for a well-defined simulation sub-domain. Within this study, we focused our attention on two packages, which play a major role in most simulation applications: geometry and electromagnetic physics.

Geometry offers the ability to describe a geometrical structure and propagate particles efficiently through it. In turn, it includes a set of sub-packages such as biasing, divisions, magnetic field, management, navigation, solids and volumes. Any simulation application involves some geometrical modeling of the experimental configuration.

The Geant4 processes package is responsible for handling particle interactions: electromagnetic interactions of leptons, photons, hadrons and ions, and hadronic interactions and transportation. Like the geometry package, it comprehends a set of sub-packages such as biasing, cuts, decay, electromagnetic, hadronic, management, optical, parameterization, scoring and transportation. Electromagnetic physics represents the core of particle transport, as almost any simulation scenario involves electromagnetic interactions either of primary or secondary particles.

\paragraph{Software Metrics Tools Selection}
For the selection of suitable software metrics tools, our criteria focused on programming language, metrics and license type. Concerning language, we found that the available tools can calculate metrics for single or multiple programming languages. Since Geant4 is in C++, we chose tools that mainly support such language. Referring to metrics, we considered tools that at least calculate code distribution, complexity, objected-oriented and size metrics. Regarding the license type, we adopted both open source tools and tools with a software evaluation license.

Our selection left us with the following tools, whose general details are shown in Table \ref{tab:gd} (where the $\surd$ symbol states that the tool has the specified detail), of which most are open source. \textbf{CCCC} (C and C++ Code Counter) \cite{cccc} analyzes C and C++ files and generates reports on various metrics of the code. \textbf{CLOC} (Count Lines of Code) \cite{cloc} counts blank lines, comment lines and physical lines of source code in many programming languages. \textbf{Pmccabe} \cite{pmccabe} calculates McCabe cyclomatic complexity and non-commented lines of code for C and C++ code. \textbf{SLOCCount} (Source Lines of Code Count) \cite{sloccount} computes SLOC in a large number of languages. \textbf{Understand} \cite{understand} is a static analysis tool for maintaining, measuring and analyzing code, supporting the main programming languages; we tried it using an evaluation license. \textbf{Unified CodeCount} \cite{unified} is a unified version of the CodeCount toolset and Difftool, supporting several languages.

\begin{table}[h]
\begin{center}
\begin{tabular}{|lrrr|} \hline
Tool & Version & Multiple Programming & Platform\\
& & Language & Independent\\
\hline
CCCC & 3.1.4 & $\surd$ & \\
CLOC & 1.60 & $\surd$ & $\surd$\\
Pmccabe & 2.6 & $\surd$ & $\surd$\\
SLOCCount & 2.26 & $\surd$ & $\surd$\\
Understand & 3.1.728 & $\surd$ & \\
Unified CodeCount & 2011.10 & $\surd$ & $\surd$ \\
\hline
\end{tabular}
\caption{General details of the selected software metrics tools.}\label{tab:gd}
\end{center}
\end{table}

\paragraph{Metrics Selection}
We considered metrics whose description is comparable on all the selected tools. Furthermore, in this initial phase we chose a minimal set of product metrics to quantify the software quality. Table \ref{tab:tm} shows the tools and metrics used in the Geant4 assessment. The $\surd$ symbol states that the tool supports the specified metrics. Our selection aggregated the product metrics in the following categories \cite{allison}:
\begin{description}
\item[Program Size (PS)] \cite{humphrey}. It includes BLOC (Blank Lines Of Code); LOC (Lines Of Code); and CLOC (Lines Of Code with Comments).
\item[Code Distribution (CD)]  \cite{humphrey}. It includes \#Files (Number Of Files); \#Methods (Number of Methods) in a class; and \#Statements (Number of Statements).
\item[Control Flow Complexity (CFC)]  \cite{watson}. It includes Traditional MCC (McCabe Cyclomatic Complexity) counting 1 for each occurrence of for, if, while, case, $\&\&$, $\mid \mid$ and ?, and adding 1 for the entire structure; and Modified MCC calculated like the Traditional MCC but counting switch rather than case.
\item[Object-Orientation (OO)] \cite{isoiec1, hyatt, henderson}. It includes CBO (Coupling Between Object classes), determining the number of other classes to which a class is coupled; DIT (Depth of Inheritance Tree), providing the maximum length from a given class to the root class of the inheritance tree; LCOM (Lack of Cohesion in Methods), defining the lack of cohesion among the methods in a class; NOC (Number of Children), determining the number of immediate sub-classes subordinated to a base class; RFC (Response For a Class), summing methods that can be executed in response to a message received by an object of the class; and WMC (Weighted Method per Class), summing the complexity of all the methods defined in the class.
\end{description}

\begin{sidewaystable}
\small
\begin{center}
\begin{tabular}{p{2cm}|ccc|ccc|cc|cccccc} 
\hline
Tool & \multicolumn{14}{c}{Metrics}\\
     & \multicolumn{3}{c}{PS} & \multicolumn{3}{c}{CD} & \multicolumn{2}{c}{CFC} & \multicolumn{6}{c}{OO}\\
 & LOC & BLOC & CLOC & \#Files & \#Methods & \#Statements & Traditional & Modified & CBO & DIT & LCOM & NOC & RFC & WMC\\
 &     &      &      &         &           &              & MCC & MCC &  &  &  & & & \\
\hline   
CCCC & $\surd$ & & $\surd$ & $\surd$ & $\surd$ & & & $\surd$ & $\surd$ & $\surd$ &  & $\surd$ & & $\surd$\\
CLOC & $\surd$ & $\surd$ & $\surd$ & $\surd$ & & &  & & & & & & & \\
Pmccabe & $\surd$ & & & $\surd$ &  & & $\surd$ & $\surd$ & & & & & & \\
SLOCCount & $\surd$ & & & $\surd$ & & & & & & & & & &\\
Understand & $\surd$ & $\surd$ & $\surd$ & $\surd$ & $\surd$ & $\surd$ & $\surd$ & $\surd$ & $\surd$ & $\surd$ & $\surd$ & $\surd$ & $\surd$ & $\surd$\\
Unified CodeCount & $\surd$ & & $\surd$ & & & $\surd$ &  &  & & & & & &\\
\hline
\end{tabular}
\caption{Tools and metrics used for the Geant4 assessment.}\label{tab:tm}
\end{center}
\small
\begin{center}
\begin{tabular}{l|cc|cc|cc|cccccc}
\hline
Maintainability  & \multicolumn{12}{c}{Metrics}\\
criteria         & \multicolumn{2}{c}{PS} & \multicolumn{2}{c}{CD} & \multicolumn{2}{c}{CFC} & \multicolumn{6}{c}{OO}\\
 & LOC & CLOC & \#Statements & \#Methods & Traditional & Modified & CBO & DIT & NOC & LCOM & RFC & WMC\\
 &     &      &      &         & MCC & MCC &  &  &  & & & \\
\hline   
Analyzability & I &I &I &I &I &I &I &I &i &I &I &I  \\
Changeability & I &I &I &I &I &I &I &I &I &I &I &I \\
Stability     & i &i &i &i &i &i &I &i &i &I &i &i \\
Testability   & I &I &I &I &I &I &I &I &i &I &I &I \\
\hline
\end{tabular}
\caption{Relationship between maintainability subcharacteristics and software metrics.}\label{tab:r}
\end{center}
\end{sidewaystable}

\paragraph{Software Quality Models Selection}
For the selection of the software quality models, we focused on maintainability, the main software characteristic as specified in the ISO 9126 standard \cite{isoiec1}. We also considered its criteria, such as analyzability, changeability, stability and testability, evaluating the effort to carry out modifications to the software.

Table \ref{tab:r} illustrates a relationship between the maintainability factor and the selected metrics \cite{lincke}: the \textit{i} letter states that the two entities typically are inversely related, the \textit{I} letter states that the inverse relation is stronger; in both cases low values of metrics are desired. We included in the ISO 9126-based model all the metrics measured from the selected tools for each Geant4 package to identify weak spots within the given release distribution.

\section{Data Collection}\label{sec: dc}

For collecting data, we installed all six tools by following the provided instructions. Some tools, such as CCCC, CLOC, Pmccabe, SLOCCount and Unified CodeCount, only provide a command-line tool; Understand provides a graphical user interface.

We downloaded various Geant4 release distributions in the range 0.0 to 10.0 in a designated directory, so that all tools were applied on the same source tree for the geometry and processes packages. Referring to the code, we performed an initial analysis on the 9.6 and 10.0 releases, since the former is widely used in production mode by several experimental applications, while the latter is the new release. It is our intention to extend the analysis to other Geant4 releases, as the historical evolution of the code could provide helpful insight for its improvement.

Concerning the tools, we used either the tool-specific export feature or the redirected operation to generate intermediate files containing the raw measured data. In most cases, the files contain unnecessary information, making it impossible to directly use them for the analysis; therefore we implemented a dedicated script both to remove and filter information before performing any evaluation.

\section{Data Evaluation}\label{sec: de}

We report some preliminary results deriving from the use of CLOC, SLOC, CCCC, Pmccabe and Understand applied to Geant4 9.6 and 10 releases. The collection of other metrics with Unified CodeCount is still in progress at the time of writing this extended abstract. Due to the large amount of metrics and tools considered, the full documentation of their values and the related analysis would be practically impossible within the limited page allocation of this paper.

Looking at some of the individual metric values per project, we observed some differences in how tools calculate these values, though they show the same trend over time. In table \ref{tab:d}, we present data showing the total, maximum, minimum and mean values for the latest two Geant4 releases. More specifically, table 4 summarizes data regarding program size and complexity in the Geant4 9.6 and 10.0 releases. Their release notes explain the LOC decreasing and the MCC reduction: removed deprecated modules and package, obsolete classes and unused methods.

\begin{table}[t]
\small
\begin{center}
\begin{tabular}{l|l|rrr|rr}
\multicolumn{7}{l}{Geant4 9.6 release (\#Files = 3897)}\\
\hline
Tools & Values & \multicolumn{3}{c}{Program Size Metrics} & \multicolumn{2}{c}{Complexity Metrics}\\
      & & BLOC & CLOC & LOC & Modified & Traditional\\
      & &      &      &     & MCC & MCC \\
\hline   
\multirow{4}{*}{CLOC} & Min & 0 & 4 & 1 & & \\
& Max & 621 & 8621 & 7379 & &  \\
& Mean & 29.48 & 62.17 & 143.31 & & \\
& Total & 114871 & 242260 & 558485 & &  \\
\hline
\multirow{4}{*}{SLOC} & Min & & & 1 & &  \\
& Max & & & 9993 & & \\
& Mean & & & 145.81& &  \\
& Total & & & 568232& &  \\
\hline
\multirow{2}{*}{CCCC} & Mean & & 66.94& 134.20& 20.02&  \\
& Total & & 260883 & 522982 & 78032&  \\
\hline
\multirow{4}{*}{Pmccabe} & Min & & & & 1&1  \\
& Max & & & & 621& 621 \\
& Mean & & & & 20.48&20.73  \\
& Total & & & & 79827&80780  \\
\hline
\multirow{4}{*}{Understand} & Min & 0& 1& 1068& 0&0  \\
& Max & 621& 3705&9036 & 403& 403\\
& Mean & 29.26& 68.58& 137.14& 17.4&17.64 \\
& Total &114031 & 267245& 534432& 67804&68755  \\
\hline
\multicolumn{7}{l}{Geant4 10.0 release (\#Files = 3629)}\\
\hline
Tools & Values & \multicolumn{3}{c}{Program Size Metrics} & \multicolumn{2}{c}{Complexity Metrics}\\
      & & BLOC & CLOC & LOC & Modified & Traditional\\
      & &      &      &     & MCC & MCC \\
\hline   
\multirow{4}{*}{CLOC} & Min & 0 & 4 & 1 & & \\
& Max & 613 & 1007 & 7379 & &  \\
& Mean & 30.44 & 59.4 & 128.21 & & \\
& Total & 110471 & 215569 & 465260 & &  \\
\hline
\multirow{4}{*}{SLOC} & Min & & & 1 & &  \\
& Max & & & 7379 & & \\
& Mean & & & 128.25& &  \\
& Total & & & 465419& &  \\
\hline
\multirow{2}{*}{CCCC} & Mean & & 61.97& 118.65& 16.20&  \\
& Total & & 224896 & 430581 & 58790&  \\
\hline
\multirow{4}{*}{Pmccabe} & Min & & & & 1&1  \\
& Max & & & & 209& 209 \\
& Mean & & & & 17.11 &17.37  \\
& Total & & & & 62082&62029  \\
\hline
\multirow{4}{*}{Understand} & Min & 0& 1& 0& 0&0  \\
& Max & 613& 1119&7378 &182& 182\\
& Mean & 30.19& 63.63& 121.32& 15.08&15.34 \\
& Total &109571 & 230911& 440270& 54711&55655  \\
\hline

\end{tabular}
\caption{Data for program size and complexity metrics in the Geant4 9.6 and 10.0 releases.}\label{tab:d}
\end{center}
\end{table}

\section{Conclusions}\label{sec: con}

With this work we aim to create a proper Geant4 data set for evaluating its software characteristics and identifying both strengths and weaknesses. In this initial study, we considered the geometry and processes packages, which play a major role in most simulation applications. The data analysis is in progress at the time of writing; its complexity is related to the differences among software metrics tools, to the large amount of data to be evaluated and to the pioneering character of this project. We are confident that the results will provide valuable indications for the Geant4 user community and constructive guidance to improve key Geant4 packages. Other Monte Carlo codes could profit from the methodology developed in this pilot project for similar assessments of their software quality.

At the end of this study, we intend to create a baseline for evaluating correlations between software quality embedded in the Geant4 development process and simulation observables produced by Geant4-based applications.

\label{------------------------------------------------END-------------------------------}

\bibliography{rpsd}
\bibliographystyle{alpha}

\end{document}